\documentstyle[osa,aps,prb,epsfig]{revtex}
\begin{document}
\twocolumn[\hsize\textwidth\columnwidth\hsize\csname
@twocolumnfalse\endcsname
\title{$^{\ast\ast}$Paramagnetic Meissner Effect in Superconducting Single Crystals of Ba$_{1-x}$K$_x$BiO$_3$}
\author{$^{\dagger\dagger}$Hyun-Tak Kim$^1$, H. Minami$^1$, W. Schmidbauer$^2$, J. W. Hodby$^2$, A. Iyo$^3$, F. Iga$^3$, and H.
Uwe$^1$}
\address{$^1$Institute of Applied Physics, University of Tsukuba, Tsukuba, Ibaraki 305,
Japan\\
$^2$Clarendon Laboratory, University of Oxford, Parks Road, Oxford
OX1 3PU, Great Britain\\
$^3$Electrotechnical Laboratory, Tsukuba, Ibaraki 305, Japan}
\maketitle{}
%\newpage
\begin{abstract}
The paramagnetic Meissner effect (PME) in the field cooled
susceptibility has been observed in superconducting single
crystals of Ba$_{1-x}$K$_x$BiO$_3$ using magnetometers of
Quantum-Design Co.. The PME is similar to that observed in
cuprates and Nb superconductors. For the
Ba$_{0.63}$K$_{0.37}$BiO$_3$ crystal, the PME is observed in
cooling process under the field up to 750 Oe. The PME was not
found in the virgin-charged superconducting magnet. The PME is
found in a superconductor with trapped flux and may arise from a
remnant field, a temperature inhomogeneity near $T_c$, and a
potassium-concentration inhomogeneity.
\\ \\
\end{abstract}
]
%\newpage
%\narrowtext
%\section{INTRODUCTION}
Recently, during a few years, paramagnetic Meissner effect (PME)
as an abnormal effect has been known that positive moment is
observed in a positive field for a superconductor. The PME has
been reported in some samples of a variety of high-$T_c$
superconductors.$^{1-7}$ It has been proposed that the effect
arose in granular hole-doped cuprates from current loops with
$\pi$ phase shift of the superconducting order parameter at some
grain boundary junction. It is argued that such behavior would be
expected to occur in a $d$-wave superconductor, but not in a
conventional $s$-wave superconductor.$^8$ The PME was found in a
Nb $s$-wave superconductor as well.$^{5,6}$ It has been insisted
that PME takes place because of a field inhomogeneity in a
magnetometer.$^{5,6}$ In this paper we report new experimental
results on the paramagnetic effect observed in the field cooling
in superconducting single crystals of Ba$_{1-x}$K$_x$BiO$_3$
(BKBO) known as the $s$-wave superconductor.

BKBO crystals, used for the observation of PME, were synthesized
by the electro-chemical method reported elsewhere.$^9$ The size of
crystal I with $T_c$=31 K was 3 x 3 x 1 mm$^3$, which was obtained
by abrading an original crystal 3.5 x 4 x 2 mm$^3$ with a sand
paper to remove insulating phase on the surface of the crystal.
The potassium concentration was analyzed as x$\approx$0.37 with
the standard deviation of 0.02 at 10 points in the $ab$ plane,
0.02 at 20 points in the $ac$ plane and 0.04 at 20 points in the
$bc$ plane by the electron-probe microanalysis. Color of the
crystal was blue up to the inside. The $H_{c1}$ at 5 K was
$\sim$750 Oe. The $\lambda$(0) of the crystal were investigated as
$\sim$5000$\AA$, respectively. In the zero-field cooled
susceptibility measured at 2 Oe in Fig. 1 (a), the sharp drop at
$T_c$ with transition with ${\triangle}T$=2 K (defined from $T_c$
to the 80$\%$ of the ZFC susceptibility at 5 K) indicates that the
superconducting phase of the crystal is nearly single and
homogeneous.

The susceptibilities were measured by using magnetometers of
Quantum-Design Co. of MPMS-5, MPMS-5S and MPMS-2 with the MOMSR2
program. Here, the magnetometers utilize a superconducting
quantum interference device (SQUID). Before measuring
susceptibilities, the magnetometer of MPMS-5 was calibrated with
the standard-sample of palladium and found to be operating
normally in its specification. The crystal was fixed in a plastic
straw in the magnetometer. Before applying a field, temperature
of a crystal was increased up to 50 K above $T_c$, and again,
decreased down to 5 K. The magnetometer paused for 5 minutes
after setting field and temperature, and then, the zero-field
cooled (ZFC) susceptibility was measured down to 5 K with
decreasing temperature from the normal state. The
susceptibilities are shown only up to 35 K in the following
figures. The field was applied from 1 to 4000 Oe. The high
resolution mode in which the magnetic field is set with a
deviation within 0.2 Oe below 5000 Oe was operated. To eliminate
most of remnant magnetic fields normally found in superconducting
magnets, magnet cleaning in the oscillating mode for
magnet-charging was carried out by dropping a field from 35,000
to 0 Oe.

Figure 1 (a)-(c) shows field dependences of FC susceptibilities
observed by MPMS-5 for crystal I. Before and after measurements,
in the scan-length mode of the magnetometer, values of X-axis in
a SQUID-response signal did not change within the scan length.
This indicates that the crystal in the straw did not move during
scan. Values of ZFC susceptibilities at 5 K are different from
each other to external fields, as shown in Fig. 1 (a), while
those of FC ones increase with decreasing the field. The
difference in the ZFC susceptibilities may be because the crystal
in the magnet is affected by the inhomogeneous field $\alpha$
before being subjected to an external field, while the decrease
of the FC ones may be because the crystal is subjected to $H_{ap}$
=$H_{ex}$ + $\alpha$ in the normal state. The paramagnetic
susceptibility decreases with increasing magnetic field and
changes to the diamagnetic one above $H_{tr}{\sim}$750 Oe. This
field $H_{tr}$ is very large compared to those found in other
crystals.$^{1-7}$ With increasing field, FC susceptibility
changes to have a minimum as a function of temperature, as shown
in Fig. 1 (c). With decreasing temperature near $T_c$ the FC
susceptibilities below 16 Oe exhibit rapid paramagnetic increase,
while the susceptibilities from 32 to 750 Oe show only a very
little diamagnetic decrease, and then, again, the paramagnetic
increase, as shown in inset of Fig. 1 (b). This phenomenon was
also observed in crystals with a large $\triangle$T and in a
relatively strong magnetic field. Furthermore, because surfaces
of crystal I were abraded as mentioned above, PME cannot be
interpreted as an impurity effect on the surface. Decrease of the
paramagnetic moment at 5 K is approximately proportional to the
logarithm of the inverse field; ${\triangle}M$ = A$ln$(1/$H$), as
shown in Fig. 2. The moment of 1 Oe may be unstable in MPMS-5.
The PME is similar to those observed in cuprate and Nb
superconductors.$^{1-7}$

Figure 3 shows FC and ZFC susceptibilities observed by MPMS-5 for
another crystal II with $T_c$=29 K. The paramagnetic
susceptibility with increasing temperature after cooling in the
field nearly coincides with one with decreasing temperature in
the field. The transition field from paramagnetic susceptibility
to diamagnetic one is $H_{tr}\approx$20 Oe. Crystal II was ground
to powder particles of sizes of 1$\mu$ order and PME was still
observed with the transition field $H_{tr}\approx$12 Oe. Thus,
PME may not arise from a temperature inhomogeneity because the
inhomogeneity might be negligibly small for particles of a size
of 1$\mu$. The decrease of $H_{tr}$ may be due to the decreased
pinning effect produced by change from the crystal to powder.

To investigate the phenomenon of the flux trap, the field
dependence of the moment was measured for the crystal I placed
parallel to the direction of the field ($H{\perp}c$-axis) with the
MPMS-5S, as shown in Fig. 4. The moment in field cooling is less
than that measured at $H{\parallel}c$-axis, which may be due to
the demagnetization effect. During measuring the FC
susceptibility at $H_{ex}$=5 Oe, the external field was dropped
to zero Oe at 5 K and then the moment was measured up to 40 K
with increasing temperature in zero field. The paramagnetic
moment increased largely. The flux was trapped. This is a
ferromagnetic phenomenon by flux trap. The field inhomogeneity as
a field variation in the scan length of 3 cm was not observed by
measuring the field by using a Hall sensor. In measuring PME, it
is independent of measurement methods, in which a sample is fixed
or moved in a magnetometer in the measurement. This was already
revealed.$^6$

Besides these two crystals shown here, PME in another dozen of
crystals was observed. The PME has been observed for crystals
with different $T_c$'s and shapes such as cylinder, disc and
random types placed perpendicular and parallel to the direction
of the field. For a high quality thin film, PME was observed. For
a crystal the temperature hysterisis phenomenon in the FC
susceptibility in the field cooling and warming cycle was found
near $T_c$; this hysterisis may occur when an intensity of the
field decreases with time. It should be noted that PME depends
upon an applied field, crystal quality, and crystal size (or
quantity).

To investigate the effect of the remnant field trapped in a
superconducting magnet, the susceptibility was measured by using
the quenched superconducting magnet. Liquid helium in the MPMS-5
was evaporated perfectly to quench the magnet, and then, the
remnant field can be ignored; the magnetic shield was imployed to
reject the external field such as earth field. The PME was not
observed with crystal I. Although it was not found, we found that
the flux was strongly trapped in the crystal because the absolute
values of FC diamagnetic susceptibilities were $\frac{1}{100}$
times as small as those of ZFC ones. Also, PME was not observed
in the virgin charge of the superconducting magnet in the MPMS-2.

The better the superconducting characteristic of crystal, the
more difficult the trap of flux is because of far less
inhomogeneity of the potassium concentration. However, if the
flux in a high quality crystal is trapped, the crystal exhibits
much higher transition field $H_{tr}$ such as crystal I. This trap
mechanism may be similar to that of a YBCO crystal with twin
boundary in a higher field than $H_{c1}$.$^{10}$ Even if the FC
susceptibility in trapped superconductors is observed as
diamagnetism, there exists evidence of the paramagnetic effect.

In conclusion, PME, which was found in the BKBO superconductor of
$s$-wave type,  cannot be interpreted as the Josephson-$\pi$
junction. The PME may arise from a remnant field trapped in a
superconducting magnet in a magnetometer, a temperature
inhomogeneity near $T_c$, and a potassium-concentration
inhomogeneity.

%\begin{center}
%\noindent{\bf ACKNOWLEDGEMENTS}
We would like to thank Dr. Kwon-Sang Ryu at KRISS in Korea for
the standardization of a Hall sensor. The Hall sensor was kindly
provided by Asahi Chemical Co..
%\end{center}

\begin{figure}
\vspace{1.0cm}
\centerline{\epsfysize=15.5cm\epsfxsize=7.5cm\epsfbox{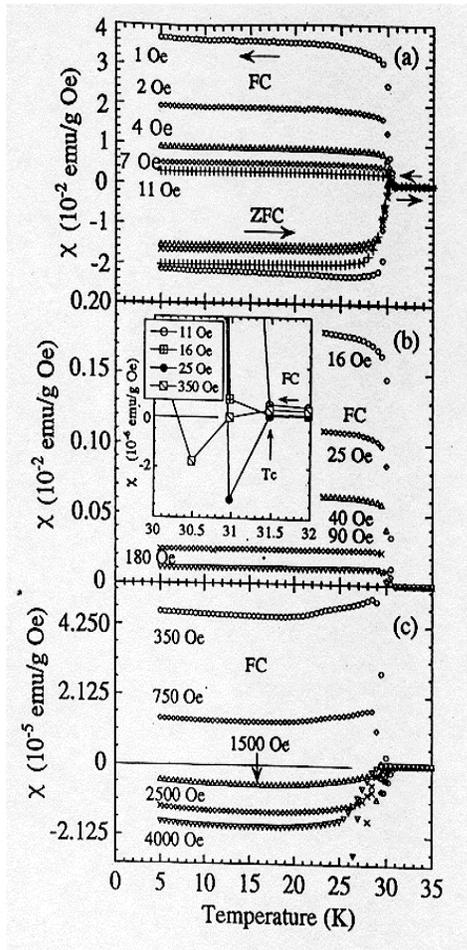}}
\vspace{0.5cm} \caption{For crystal I, (a) ZFC and FC
susceptibilities at external fields of 1 to 11 Oe, (b) FC
susceptibilities at external fields of 16 to 180 Oe, and (c) FC
susceptibilities at external fields of 350 to 4000 Oe. Inset of
Fig. 1 (b), FC susceptibilities near $T_c$ at external fields of
11 to 350 Oe for crystal I.}
\end{figure}

\begin{figure}
\vspace{2.0cm}
\centerline{\epsfysize=9.0cm\epsfxsize=7.0cm\epsfbox{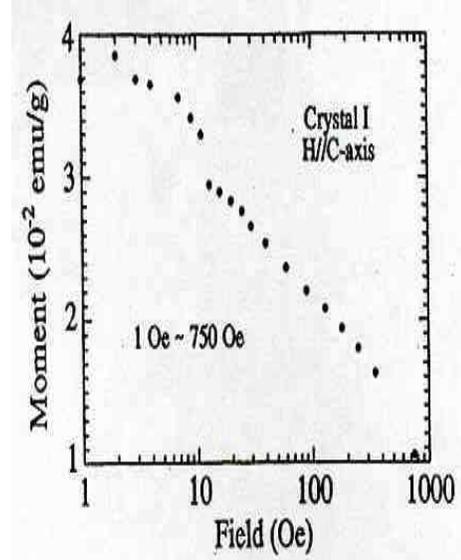}}
\vspace{0.5cm} \caption{Paramagnetic moment vs magnetic field at 5
K for crystal I.}
\end{figure}

\begin{figure}
\vspace{-1.0cm}
\centerline{\epsfysize=10.0cm\epsfxsize=7.0cm\epsfbox{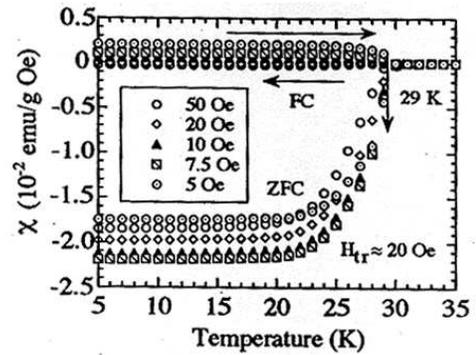}}
\vspace{-2.0cm} \caption{Field dependence of ZFC and FC
susceptibilities for crystal II.}
\end{figure}

\begin{figure}
\vspace{1.0cm}
\centerline{\epsfysize=10.0cm\epsfxsize=7.0cm\epsfbox{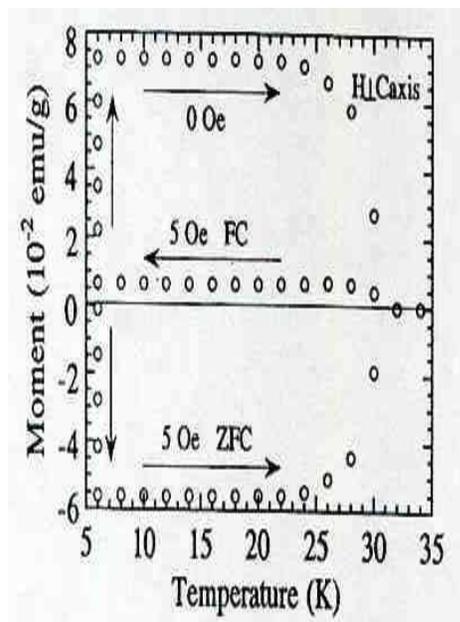}}
\vspace{0.5cm} \caption{Moment vs temperature for crystal I placed
the $H{\perp}c$-axis.}
\end{figure}

\end{document}